\documentclass[11pt,a4paper]{article}
\pdfoutput=1
\usepackage{jheppub}

\usepackage{amsmath,amssymb,latexsym,slashed,multirow,bm}

\def\tr{{\rm Tr\,}}
\def\det{{\rm det\,}}
\def\lmatrix{\left(\begin{array}}
\def\rmatrix{\end{array}\right)}
\def\bea{\begin{eqnarray}}
\def\eea{\end{eqnarray}}
\def\nn{\nonumber}
\def\msbar{\overline{\rm MS\kern-0.5pt}\kern0.5pt}
\def\gsim{\mathrel{\rlap{\lower4pt\hbox{\hskip1pt$\sim$}}\raise1pt\hbox{$>$}}}
\def\lsim{\mathrel{\rlap{\lower4pt\hbox{\hskip1pt$\sim$}}\raise1pt\hbox{$<$}}}
\def\rho{\varrho}

\title{QFT on rotating boxes at finite temperature}

\author{Sebestyen Nagy, }
\author{Daniel Nogradi}

\affiliation{Department of Theoretical Physics, Eotvos University \\ Pazmany Peter setany 1/a, Budapest 1117, Hungary}

\emailAdd{senbeastgyen@student.elte.hu}
\emailAdd{nogradi@bodri.elte.hu}

\abstract
{
We formulate thermal quantum field theory on a finite spatial periodic volume incorporating finite
rotations.
Traditional compactifications at finite temperature without rotations
typically involve ${\mathbb T}^4$ as the space-time manifold within a path integral formulation 
and also moving frames can be
accommodated by shifted boundary conditions on the same space. We show that consistent
descriptions of certain finite rotations are possible on space-time manifolds with topology
different from ${\mathbb T}^4$ but still flat and without boundary and we classify all possible geometries. 
The non-trivial
topology may be implemented by rotated boundary conditions allowing for a path integral formulation.
The purely imaginary 
angular velocity in temperature units cannot be arbitrary but several discrete values are possible. 
We also discuss finite volume effects in detail.
}


\arxivnumber{2509.19933}

\begin{document}

\maketitle

\section{Introduction\label{introduction}}

The most successful non-perturbative description of quantum chromodynamics (QCD) is via a path integral
formulation. This approach allows the calculation of various properties of quarks and gluons at finite
volume, finite temperature and coupling to external parameters such as conserved charges and
background fields. Matter made up of quarks and gluons may also rotate at a constant angular velocity
which may have significant effect on the phase diagram and lead to novel phenomena such as the 
chiral vortical effect and others
\cite{Vilenkin:1979ui,Erdmenger:2008rm,Banerjee:2008th,Son:2009tf,Kharzeev:2010gr,Landsteiner:2011cp,Kharzeev:2015znc,
Chen:2015hfc,Ebihara:2016fwa,Fukushima:2018grm,Fukushima:2020ncb,Chen:2022smf,Chernodub:2022veq,Chernodub:2022qlz,
Chen:2024tkr,Fukushima:2025hmh}.

In several experimental and observational
scenarios rotation of QCD matter cannot be ignored. For instance in fast rotating neutron stars 
\cite{Grenier:2015pya,Watts:2016uzu,Paschalidis:2016vmz} and in relativistic 
off-central heavy ion collisions \cite{Becattini:2007sr,Jiang:2016woz}. Naturally, there are many orders
of magnitude differences in angular velocities between the two scenarios. In the latter case, which will
be the focus of this paper, the experimental findings on $\Lambda$ hyperon spin polarizations indicate 
vorticity of the plasma may reach as high as about $\sim 10\, MeV$ \cite{STAR:2017ckg}.

The most natural approach to the spatial geometry
would involve a finite cylinder. The radius perpendicular to the axis of rotation would necessarily be
finite. This geometry, naturally, allows for
arbitrary infinitesimal velocities, with an upper bound coming from the fact that in a relativistic 
theory the edge of the cylinder should not be moving faster than the speed of light. Nevertheless there
is a well-defined infinitesimal rotational generator which commutes with the Hamiltonian. In quantum field theory,
the simplest implementation would involve the introduction of a rotating frame
\cite{Ambrus:2014uqa, Chernodub:2016kxh}, leading to a non-trivial space
dependent metric tensor. The non-trivial metric on the cylindrical geometry presents, however, a crucial
problem, namely it complicates the non-perturbative formulation on the lattice. First of all, it is not clear
what the right boundary conditions are spatially, second, the lattice discretization itself 
is not compatible with infinitesimal rotations which feeds into the first problem of spatial boundary
conditions and third, the definition of temperature in a curved background is non-trivial.

The approach has nevertheless been used in lattice simulations 
\cite{Yamamoto:2013zwa, Chernodub:2017ref, Braguta:2021jgn, Braguta:2023iyx} with rather surprising results
involving negative moment of inertia in a certain temperature range \cite{Braguta:2023yjn}.

Given the importance of phenomena related to rotations and their 
non-perturbative lattice studies here we propose a different approach. In the
continuum the spatial box will be a periodic 3-torus just as the starting point of most lattice studies 
without rotation. Even though the 3-torus does break Poincare invariance,
depending on its geometry a finite subgroup of the Poincare group can be preserved. 
Since there are not any infinitesimal rotational generators, it is not possible to include the effect 
of rotation in a non-trivial metric. 
Instead, after Wick rotation, the unitary operator corresponding to the preserved finite 
rotation over the full period in Euclidean time, will be inserted into the partition function directly. 
This will then be represented as a path integral over fields subject to novel rotated boundary conditions in the
Euclidean time direction. Discretization does not then present conceptual problems, 
no further breaking of the preserved
rotational group occurs by a hypercubic lattice. The rotated temporal boundary conditions are
straightforward to implement with a well-defined continuum limit at finite volume and temperature.

Technically, the continuum space-time manifolds we propose are closely related to the 4-torus, ${\mathbb
T}^4$, most frequently used at finite volume and temperature. They are all also compact, flat, orientable
and without boundary and are of the form ${\mathbb T}^4 / G$ with a finite group $G$ that involves the
aforementioned unbroken rotational subgroup of $SO(3)$. The topology of space-time with a non-trivial $G$
will be different from ${\mathbb T}^4$ though.

It is perhaps useful to draw a parallel with the introduction of other external parameters, e.g. the quark
chemical potential. In a path integral formulation there are at least two approaches. One, the action is
modified locally by the chemical potential whereas the fields remain periodic, or two, 
leaving the action the same as before, but introducing the chemical
potential in a boundary condition over the full period in the temporal direction. In the case of
rotations the introduction of the non-trivial metric locally corresponds to the first choice and thereby
the action has explicit dependence on the velocity. This is however not possible to do, even in the
continuum, on a 4-torus. With a different geometry, like a cylinder, the continuum setup would be
meaningful but a lattice discretization will introduce more breaking. Our approach is akin to the second
setup whereby the entire effect is encoded in a temporal boundary condition and the action is the same as
without rotation. In this way no mention is made of infinitesimal rotations thereby allowing for a
well-defined discretized path integral formulation respecting the finite rotations of the continuum
setup.

Needless to say, regardless of which of the two approaches one takes in a path integral formulation, all
physical quantities will be the same, provided both approaches can be defined. 
Concretely, whether rotation is introduced by modifying locally the
Lagrangian by an angular velocity dependent metric (corresponding to a rotating frame) and keeping the
same boundary conditions on the fields as without rotation (the perhaps more traditional approach), or
two, by keeping the Lagrangian the same as without rotations, but modifying the boundary conditions on
the fields in an angular velocity dependent way (as in our work), the partition function and all other
quantities will be the same. In order for both approaches to make sense, of course, infinitesimal
rotations should be compatible with the spatial volume. If this is not the case, as in our work with a
spatial torus, only the latter approach makes sense and only for those finite rotations which leave it
invariant.  The presence of rotations in this case will not be identified by a non-vanishing vacuum
expectation value for a local angular momentum operator (which is an infinitesimal generator), but rather
by the non-unit value for the vacuum expectation value of the corresponding unitary operator. Even though
from the logarithm of the unitary operator an angular momentum operator can be formally defined, this
will not be local. Nevertheless, as we will see, a consistent description can be given by using finite
rotations only, without reference to local infinitesimal generators. 

\section{Conserved charges\label{movingandrotatingboxes}}

In order to discuss finite volume and finite temperature let us assume space-time is compactified in such
a way that it remains flat, orientable and without boundary. 
Later we will see that the most often used manifold, the 4-torus, is not the only choice topologically.

The partition function in the presence of a number of local conserved charges $Q_A$ and chemical potentials
$\alpha_A$ can be written as,  
\bea
\label{z}
Z(\beta,\alpha_A) = \tr e^{- \beta \left( H - \alpha_A Q_A \right)}\;,
\eea
where $\beta = 1/T$ is the inverse temperature and $H$ the Hamiltonian, 
which is assumed to be Poincare invariant.
For instance, since the total spatial momentum $P_k$ is conserved, it can be chosen with
velocities as parameters, inserting $\exp(\beta v_k P_k)$ into the partition function trace. 
This setup corresponds to
studying the system in a moving frame. In a Euclidean setup the velocity is naturally chosen purely imaginary, $i v_k =
T y_k$ with a real spatial vector $y_k$. The partition function in this case can be computed via a
path integral formulation where the fields satisfy shifted boundary conditions
\cite{Giusti:2010bb,Giusti:2011kt,Giusti:2012yj,Giusti:2014ila,Giusti:2016iqr},
\bea
\phi\left(t + L_0, {\boldsymbol x} \right) = \phi( t, {\boldsymbol x - \boldsymbol y})\;,
\eea
where $L_0$ is the temporal extent and $\sqrt{L_0^2 + {\boldsymbol y}^2} = 1/T$.
It is worthwhile to point out that shifted boundary conditions are implemented on ${\mathbb T}^4$, 
topologically the same manifold on which one would formulate the theory without shifts i.e. zero velocity.

The analogous setup for a rotating frame would start with (\ref{z}), the role of the chemical potentials
would be played by angular velocities and the charges would be the angular momentum operators $J_k$. After
Wick-rotation the angular velocities, $\omega_k$ become purely imaginary, $i\omega_k = T \vartheta_k$, with
real angles $\vartheta_k$, leading to the insertion of $\exp(i \vartheta_k J_k)$.

The fundamental problem is that the rigid rotation of an infinite spatial volume is incompatible with a
relativistically invariant theory. At the same time, once the spatial volume is compact, flat, orientable
and without boundary, the full 
rotation group $SO(3)$ is broken so there are no infinitesimal generators $J_k$ as there were in infinite volume. 
Hence the need for a formulation without reference to infinitesimal rotational generators.

The key observation is that a finite subgroup of $SO(3)$ can
be maintained with corresponding unitary operators $U(\vartheta)$ leading to the partition
function,
\bea
\label{zzz}
Z(\beta,\vartheta) = \tr e^{- \beta H} U(\vartheta)\;,
\eea
which, as we will see, can again be represented as a path integral over fields subject to novel
boundary conditions. The finite angles will not be arbitrary however. The simplest variant will be,
\bea
\label{simple}
\phi\left(t + \frac{1}{T}, {\boldsymbol x} \right) = \phi( t,  {\boldsymbol { R^{-1} x } } )\;,
\eea
with ${\boldsymbol R} \in SO(3)$, and periodicity spatially. More complicated
boundary conditions are possible with periodicity up to rotations in the spatial directions as well.
One of the
interesting aspect of these new boundary conditions is that they correspond to fields living not on
${\mathbb T}^4$ but topologically distinct but still flat, compact and orientable manifolds. There are
only finitely many topologically different such manifolds and we will classify the possible geometries
(metrics). Furthermore,
rotation and movement simultaneously is also possible, provided the rotations commute with the
corresponding translations. Thus the most general setup we will be discussing relates to the Euclidean
partition function,
\bea
\label{rotmove}
Z(\beta, \vartheta, y) = \tr \left[ e^{-\beta H} e^{i y_k P_k} U(\vartheta) \right],
\eea
with some choices of angles and translations $y$. Expressed succinctly, the goal of our paper is to 
formulate Euclidean QFT defined by (\ref{rotmove}) via a path integral on a flat, compact space-time
without boundary.

\section{QFT on compact, flat, orientable manifolds\label{qftoncompact}}

After introducing our main motivation, rotating frames at finite temperature and finite volume, let us
generalize the setup somewhat. We are interested in compactifications of space-time ${\mathbb R}^4$ with
the standard metric, which
remain flat so locally cannot be distinguished from ${\mathbb R}^4$, and are orientable and of
course compact. The prototypical example is ${\mathbb T}^4$. 
It turns out all others can be realized as ${\mathbb T}^4 / G$ where $G$ is a finite subgroup of the Euclidean group
consisting of translations and rotations \cite{bieberbach1, bieberbach2}; see 
\cite{hatcher, Ratcliffe:2006bfa, book} for more
details. More
precisely, let us start with a set of basis vectors in ${\mathbb R}^4$, $e^{a}$ assembled into a
$4\times 4$ matrix $e^{a}_\nu$, the $a^{th}$ column is $e^a$. There is an
associated lattice $\Lambda \subset {\mathbb R}^4$ given by integer linear combinations $e^{(n)} = n_a
e^{a}$. Then we have ${\mathbb T}^4 = {\mathbb R}^4 / \Lambda$ and the volume of this 4-torus is
$V = |\det(e)|$. There are stringent conditions on subgroups $G$ of the Euclidean orientation preserving
group $SO(4) \ltimes {\mathbb
R}^4$ such that they descend to an action on ${\mathbb T}^4$ in a way that ${\mathbb T}^4 / G$ is a
smooth manifold without boundary. 
If elements of $G$ are written as $g = (A,b)$ with a rotation $A$ and translation $b$
acting as $g\cdot x = Ax+b$, then
for instance, $A$ should preserve the lattice $\Lambda$, i.e. its matrix elements in the basis of
$e^{a}$ should all be integers $N_{ac} \in {\mathbb Z}$ so 
$A e^{c} = N_{ac} e^{a}$. Furthermore, the translations
$b$ should be such that the action on $G$ is without fixed points. In the $e^{a}$ basis the
translations $b$ have rational coefficients, $b = r_a e^a$ with $r_a = k_a / |G|$ and integer $k_a$. 
These conditions restrict the
possibilities for $G$ and allow for 26 distinct cases different from ${\mathbb T}^4$, 
meaning there are 26 non-trivial compact, flat, orientable 4-manifolds with different
topologies. On each one of these manifolds the metric descends from that of the original ${\mathbb T}^4$,
which in turn descended from ${\mathbb R}^4$ and
is fully determined by $e$, and can be parametrized by a number of continuous parameters. Clearly, if
$R\in SO(4)$ and $K \in GL(4,{\mathbb Z})$ then $e$ and $R\, e\, K^{-1}$ correspond to the same 4-torus. Here
the freedom with respect to $R$ is just an over-all orientation of all 4 basis vectors $e^a$, while $K$
corresponds to a relabeling of the integer vectors $n_a$ and hence does not change the lattice $\Lambda$
and consequently the resulting ${\mathbb T}^4$. We will use these transformations to reduce $e$ to the 
simplest possible form.

It is sometimes useful to realize that for our flat, compact and orientable manifolds we have two
realizations, ${\mathbb T}^4 / G \sim {\mathbb R}^4 / ( G \ltimes \Lambda )$.  It is equivalent to first 
factor by the lattice $\Lambda$ and end up with ${\mathbb T}^4$ and further factoring by $G$, or have a single
factorization of ${\mathbb R}^4$ by $G \ltimes \Lambda$. With this in mind let us introduce $G_\Lambda =
G \ltimes \Lambda$, its elements are $(A,b,e^{(n)})$. Clearly, 
$G$-invariant functions on ${\mathbb T}^4$, which are well-defined on ${\mathbb T}^4/G$, 
can be defined as $G_\Lambda$-invariant functions on ${\mathbb R}^4$. 
The topology of
${\mathbb T}^4/G$ is completely determined by $G_\Lambda$, in particular the fundamental group of the former is
the latter, $\pi_1({\mathbb T}^4/G) = G_\Lambda$.

Once a flat, compact and orientable ${\mathbb T}^4 / G$ is given let us turn to the Fourier transform
with a view towards QFT.
A basis for the reciprocal lattice $\Lambda^*$ will be denoted by $f^{a}$, i.e. $(f^{a},e^{b}) =
\delta_{ab}$. On ${\mathbb T}^4$ the Fourier modes are $e^{i p x}$ with $p = 2\pi n_a f^{a} =
2\pi f^{(n)}$ with
some integers $n_a$, as usual. The following Fourier modes are symmetric with respect to $G$ and hence
are well-defined on ${\mathbb T}^4/G$,
\bea
\phi_p(x) = \frac{1}{|G|} \sum_{(A,b)\in G} e^{i p (Ax+b)}\;.
\eea
Not all modes are independent, 
if $q = Ap$, with some $(A,b)\in G$ then $\phi_p(x)$ and $\phi_q(x)$ are related by,
\bea
\phi_p(x) = e^{-i q b} \phi_q(x)\;,
\eea
hence momenta $p$ and $Ap$ should be regarded as equivalent. Furthermore, if the momenta is such 
that $p = Ap$ for all $(A,b)\in G$, some of the corresponding modes may be identically zero. How many
such modes, if at all, exist depends non-trivially on $G$. In the simplest case of an abelian group, the
type considered in this paper in detail, the momenta corresponding to vanishing modes, i.e. ``forbidden''
momenta can be characterized as follows. There is a single generator $(A,b)$ and momenta left invariant
by it can be labelled by two integers corresponding to the two directions left invariant by $A$. One of
these directions is $b$ and if the corresponding integer is not a multiple of $|G|$, these momenta are
``forbidden''. More details will be given in a forthcoming publication. 

\subsection{Free scalars}

As a simple example let us consider free scalars.
The Fourier modes $\phi_p(x)$ are eigenmodes of the Laplacian on ${\mathbb T}^4/G$, $-\Delta \phi_p(x) = p^2 \phi_p(x)$
because all $A$ are orthogonal. The only difference relative to ${\mathbb T}^4$ is that the multiplicity
of each eigenvalue $\lambda(n) = p^2 = 4\pi^2(f^{(n)},f^{(n)})$ given in terms of the integer vector
$n_a$ can be less than on ${\mathbb T}^4$. This is
because $p$ and all $Ap$ with $(A,b)\in G$ are identified and should be counted only once. 

Let us calculate the partition function $Z = e^{-Vf}$ and free energy density $f$
for a free massive scalar on ${\mathbb T}^4/G$.
The methods are standard and use sums over $G_\Lambda$ of the propagator (also called mirror charges),
instead of directly the regularized product of eigenvalues $\lambda(n) + m^2$.
The free energy density first on ${\mathbb T}^4$ is made finite by substracting a divergent contribution
corresponding to the infinite volume result. The finite part may be obtained from
the propagator on ${\mathbb R}^4$,
\bea
\label{prop}
G_{{\mathbb R}^4}(x,y) = \frac{m^2}{4\pi^2} \frac{K_1(m|x-y|)}{m|x-y|}.
\eea
Summing over $\Lambda$, except $n=0$, corresponding to the subtraction we get, after Poisson
resummation \cite{Giusti:2012yj}, 
\bea
\Delta(x) &=&  \frac{1}{|x|}\frac{\partial}{\partial |x|} G_{{\mathbb R}^4}(x,0) \nn \\
f_{{\mathbb T}^4} &=&  \sum_{ n \neq 0} \Delta( e^{(n)} ).
\eea
Now turning to ${\mathbb T}^4/G$, the partition function can again be made finite by the subtraction of
the same term and we obtain the manifestly finite result,
\bea
\label{part}
&& f_{{\mathbb T}^4/G} = \\
&=& \sum_{\footnotesize{ \begin{array}{c} (A,b,e^{(n)}) \in G_\Lambda \\ (A,b) \neq (1,0) \end{array}}} \Delta(b + e^{(n)}) +
\sum_{n\neq 0} \Delta( e^{(n)} ), \nn
\eea
which determines the free energy density explicitly in terms of the metric on ${\mathbb T}^4$ encoded in $e$, the
finite group $G$ and the mass $m$. The infinite sums in (\ref{part}) are fast converging thanks to the
exponential decay of the Bessel function in $\Delta(x)$ for large arguments.

\section{Rotation and finite volume effects}

Out of the 26 non-trivial flat, compact, orientable spaces a subset will have an interpretation of 
finite rotations at finite temperature. Some of the corresponding groups $G$ have a single generator and
hence abelian and some of them have two generators, either abelian or non-abelian. In this paper we will
only deal with the cases of a single generator in detail and will only briefly mention the setups 
corresponding to the potentially non-abelian cases.

The lattices $\Lambda$, with the interpretation of finite rotations at finite temperature,
are defined by the basis vectors of the form,
\bea
\label{eE}
e = \lmatrix{cc} L_0 & 0 \\ 0 & E \rmatrix,
\eea
with the Euclidean time and spatial directions orthogonal, and correspondingly we require that the 
generator $(A,b) \in G$ acts via a spatial rotation by $A$ and temporal shift by $b$.
Here $E$ is a $3\times 3$ matrix characterizing
the metric on the spatial ${\mathbb T}^3$, similarly to our description of $e$ and ${\mathbb T}^4$. The
columns are denoted by $E^a$ but of course now $a$ runs from 1 to 3 only and $E^{(n)} = n_a E^a$. The
spatial volume is $V_3 = |\det(E)|$.

Before considering the suitable groups $G$ let us briefly describe the 3-torus defined by $E$. This is
also relevant at zero temperature simply because in the zero temperature limit
our space-time is ${\mathbb R} \times {\mathbb T}^3$. Hence finite volume effects on masses and decay
constants at zero temperature are determined by the geometry of ${\mathbb T}^3$ i.e. the matrix $E$.

In the most frequently used case we have $E = {\rm diag}(L_1,L_2,L_3)$ and finite volume effects in QFT 
on such an orthogonal 3-torus are well-understood \cite{Luscher:1985dn}. The masses of stable particles
at finite volume depend on the infinite volume mass $m$ as well as the spatial sizes. In
the absence of a 3-point vertex the leading finite volume corrections, $\delta m$, are given in terms of the 
infinite volume propagator (\ref{prop}), 
\bea
\label{delta}
m(L_1,L_2,L_3) &=& m + \delta m\\
\delta m &=& \frac{\lambda}{4m} \sum_{n \neq 0} G_{{\mathbb R}^4}(E^{(n)},0), \nn
\eea
i.e. the propagator is evaluated at $x_0 = y_0$ and the separations spatially are all the possible paths,
$E^{(n)}$,
along which the particle can loop around the finite volume. Here $\lambda$ is the 4-point vertex, assumed to be a
constant, as is the case for instance at leading order in chiral perturbation theory, $\lambda = 2 m^2 / ( f^2 N_f )$
with $N_f$ flavors and decay constant $f$. 

The particular form of the 4-point vertex will not be relevant for us, only the dimensionless factor,
\bea
\label{delta2}
\delta(E) = \frac{4\pi^2}{m^2} \sum_{n \neq 0} G_{{\mathbb R}^4}(E^{(n)},0)
\eea
which will be the focus from now on. The asymptotic form of the Bessel-function ensures an
exponentially small correction and the largest contribution is coming from the smallest exponent. 
This in turn, in $m$ units, is the smallest distance between 
the origin and points in the spatial lattice, due to the sum in (\ref{delta2}). 
If the goal is to minimize finite volume effects, at a fixed $V_3 =
|\det(E)|$, then the task is to find $E$ such that this smallest distance is maximal. With an
ortogonal 3-torus this clearly occurs if $L_1 = L_2 = L_3 = L$ in which case we end up with
the standard result \cite{Gasser:1986vb},
\bea
\label{deltat3}
\delta(L) = 6 \frac{K_1(mL)}{mL} + 12 \frac{K_1(\sqrt{2} mL)}{\sqrt{2}mL} + \ldots. 
\eea

Let us generalize the above for a 3-torus with non-orthogonal basis vectors $E$. In
this case fields are periodic spatially in a generalized sense, 
$\phi(t,{\boldsymbol x}+E^a) = \phi( t, {\boldsymbol x})$ for $a=1,2,3$. 
This simply means that instead of an orthogonal (rectangular) parallelepiped they are defined on a general
parallelepiped and the faces across from each other are identified. Finite
volume effects are still determined by (\ref{delta2}) but the search for smallest corrections leads to the
following non-trivial optimization problem: one would like to find, for each spatial lattice, i.e. fixed
$E$, the minimal distance between lattice points and the origin and one would like to maximize this minimal
distance over all lattices. All of this at a prescribed fixed volume $V_3$. 

It turns out the solution to the problem is known from the study of 3-dimensional sphere packings 
and is given by the following basis vectors,
\bea
\label{optimale}
E = \frac{L}{2^{1/3}} \lmatrix{ccc} -1 & 0 & 1 \\ 1 & 1 & 1 \\ 0 & 1 & 0 \rmatrix,
\eea
where the normalization is such that $V_3 = L^3$. Clearly, the vectors $E^a$ are not all orthogonal. 
The corresponding spatial lattice may be characterized as the set of points
$(x_1,x_2,x_3)$ such that the sum $x_1 + x_2 + x_3$ is even in units of $L/2^{1/3}$; also
called a tetrahedral-octahedral honeycomb lattice. It is easy to see
that finite volume effects are in fact exponentially smaller on this 3-torus than on the
standard orthogonal 3-torus of equal lengths. In the latter case we have (\ref{deltat3}) while with the
former, using (\ref{delta2}),
\bea
\label{optim}
\delta(L) = 12 \frac{K_1(2^{1/6} mL)}{2^{1/6} mL} + 6 \frac{K_1(2^{2/3} mL)}{2^{2/3}mL} + \ldots\;\;\;
\eea
The leading exponent $2^{1/6} = 1.1225...$ is larger than $1$ and the subleading exponent
$2^{2/3} = 1.5874...$ is also larger by the same factor than the corresponding subleading 
exponent in (\ref{deltat3}) which is $\sqrt{2} = 1.4142...$.

By various choices of $E$, in other words various choices of metrics on ${\mathbb T}^3$, 
it is possible to obtain leading exponents $\alpha$ anywhere in the range
$1 \leq \alpha \leq 2^{1/6}$. The two cases $\alpha = 1$ and $\alpha = 2^{1/6}$ are 
clearly of special interest, the former for corresponding to the most frequently used case and the latter
for corresponding to the smallest finite volume effects. 

Once ${\mathbb T}^4$ is given and its metric defined by (\ref{eE}) with some $E$ let us describe the
various allowed groups $G$. As mentioned before, here we will only consider abelian groups with a single
generator $g = (A,b)$ such that $A$ rotates spatially and $b$ is a temporal shift. 
Without loss of generality these may be taken as,
\bea
\label{gen}
A &=& \lmatrix{cccc} 
1 & 0 & 0 & 0 \\
0 & \cos(2\pi/k) & -\sin(2\pi/k) & 0 \\ 
0 & \sin(2\pi/k) & \;\;\;\cos(2\pi/k) & 0 \\ 
0 & 0 & 0 & 1 
\rmatrix \nn \\
b &=& \lmatrix{c} L_0 / k \\ 0 \\ 0 \\ 0 \rmatrix  
\eea
with $k=2,3,4,6$ and as we will see there will be two choices for $k=2,3,4$ and a unique choice for
$k=6$. No other integer $k$ is possible.
The two families will be denoted by $G = {\mathbb Z}_k$ with $k=2,3,4,6$ and $G = {\mathbb
Z}_k^\prime$ with $k=2,3,4$. Hence all together we will discuss 7 cases. Clearly $g^k = 1$. 
The spatial rotations
$\boldsymbol R$ are in the $x-y$ plane by angle $\vartheta = 2\pi/k$ and the shift by $b$ is indeed
in the temporal direction. The space-time volume of the resulting space ${\mathbb T}^4 / G$ is 
$V / k$ where $V$ is the space-volume of the original ${\mathbb T}^4$.
The boundary conditions for fields on ${\mathbb T}^4 / G$ are then,
\bea
\label{rrr}
\phi\left(t+\frac{L_0}{k},x_1,x_2,x_3\right) & = & \phi(t,x_1^\prime,x_2^\prime,x_3) \nn \\
\phi\left(t,{\boldsymbol x} + E^a\right) & = & \phi(t,{\boldsymbol x})
\eea
where,
\bea
\label{xprime}
x_1^\prime &=& \;\;\; x_1\cos(2\pi/k)+x_2\sin(2\pi/k)  \\
x_2^\prime &=& -x_1\sin(2\pi/k)+x_2\cos(2\pi/k) \nn
\eea
Now setting $L_0/k = 1/T$ we see that these are precisely the boundary conditions for a
path integral representation of (\ref{zzz}) in the form (\ref{simple}). 

Let us describe the possible metrics given by $E$, compatible with the above action of $G$ on 
${\mathbb T}^4$ such that ${\mathbb T}^4/G$ is a smooth manifold. We will see that for a given $G$
the allowed metrics $E$ will contain a number of free parameters. These may or may not contain the
orthonormal special case. Also, the search for the smallest finite volume effects, i.e. search for largest
$\alpha$, will result in $\alpha_{max}$ which may or may not equal the largest possible value $2^{1/6}$.
This is because for given $G$, the form of $E$ will be restricted relative to the most general case, and
hence the optimal solution for $\alpha_{max}$ will necessarily be $\alpha_{max} \leq 2^{1/6}$, but in some
cases, as we will see, strictly less.

In each case $E$ is normalized such that $V_3 = L^3$ and the shortest path between lattice points and
the origin is computed by the Lenstra-Lenstra-Lov\'asz algorithm \cite{Lenstra:1982eee}.

\section{Metrics on ${\mathbb T}^4$}

Once the coordinates are chosen in all 7 cases such that the generators $g = (A,b)$ are given by
(\ref{gen}), the most general metrics, compatible with the action of $G$ on ${\mathbb T}^4$, are listed
below, together with the optimal exponents $\alpha_{max}$ corresponding to finite volume effects.

\subsection{${\mathbb Z}_2$}

The metric on ${\mathbb T}^4$, encoded in $e$ may be chosen in the form (\ref{eE}) with,
\bea
\label{z2E}
E = 
 \frac{L}{|z_2 z_3|^{1/3}} \lmatrix{ccc}
 1 & 0 & 0 \\
 z_1 & z_2 & 0 \\
 0 & 0 & z_3 \rmatrix,
\eea
where $z_1$ is arbitrary and $z_{2,3}\neq 0$. An orthonormal 3-torus is possible with
$(z_1,z_2,z_3)=(0,1,1)$, while the smallest finite volume effects occur at $(z_1,z_2,z_3) =
\left(1/\sqrt{3},2/\sqrt{3},2/\sqrt{3}\right)$ leading to $\alpha_{max} = (4/3)^{1/6} = 1.0491...$, our first example
where the finite group $G$ is not compatible with a 3-torus of absolute smallest finite volume effects,
$\alpha_{max} < 2^{1/6}$.

\subsection{${\mathbb Z}_3$}

In this case we have,
\bea
\label{z3}
E =
 L \left( \frac{\sqrt{3}}{2|z|} \right)^{1/3} \lmatrix{ccc} 1 & 0 & 0 \\
 \frac{1}{\sqrt{3}} & \frac{2}{\sqrt{3}} & 0 \\
 0 & 0 & z \rmatrix,
\eea
with $z\neq 0$. The previous 3-torus (\ref{z2E}) is a special case with $(z_1,z_2,z_3) =
(1/\sqrt{3},2/\sqrt{3},z)$. Clearly, an orthonormal 3-torus is not possible with any choice of the free
parameter $z$, although $|E^1| = |E^2|$, 
the angle between $E^1$ and $E^2$ is $\pi/3$. The smallest finite volume effects occur at
$z = 2/\sqrt{3}$ and we have $\alpha_{max} = (4/3)^{1/6}$ again.

\subsection{${\mathbb Z}_4$}

With $G = {\mathbb Z}_4$ we have
\bea
E = 
\frac{L}{|z|^{1/3}} \lmatrix{ccc} 1 & 0 & 0 \\
0 & 1 & 0 \\
0 & 0 & z \rmatrix,
\eea
with $z\neq 0$ and clearly the corresponding 3-torus is always orthogonal. The choice $z=1$ corresponds
to the orthonormal setup which is also the one with the smallest finite volume effects, $\alpha_{max}=1$, 
within this family of 3-tori. Again, $\alpha_{max} < 2^{1/6}$, strictly less than the over-all optimal.

\subsection{${\mathbb Z}_6$}

The allowed metrics and all properties are identical to that of ${\mathbb Z}_3$, see (\ref{z3}). This
means that the same spatial 3-tori which are compatible with $2\pi/3$ rotations are also compatible with
$\pi/3$ rotations as well.

\subsection{${\mathbb Z}_2^\prime$}

In this case we have,
\bea
E = \frac{L}{(2|z_2z_3|)^{1/3}} \lmatrix{ccc} 
z_1 & - z_1 & z_3 \\
z_2 & - z_2 & 0     \\
1     &   1     & 0    
\rmatrix, 
\eea
with arbitrary $z_1$ and $z_{2,3}\neq0$, so $|E^1| = |E^2|$ while $|E^3|$ is arbitrary.
The choice $(z_1,z_2,z_3) = (0,1,1)$ leads to an orthonormal 3-torus while 
$(z_1,z_2,z_3) = (\sqrt{2},1,2\sqrt{2})$ corresponds to the smallest finite volume effects with
$\alpha_{max} = 2^{1/6}$, the largest possible value.
The latter one is seen to be equivalent to (\ref{optimale}) by an over-all rotation $SO(3)$ and 
$GL(3,{\mathbb Z})$ transformations.

\subsection{${\mathbb Z}_3^\prime$}

The second topology compatible with $\vartheta = 2\pi/3$ rotations is given by the spatial metric,
\bea
E = L \frac{2^{1/3}}{|z|^{1/3} \sqrt{3}} \lmatrix{ccc} 
 \frac{1}{2} & \frac{1}{2}  & 1 \\
-\frac{\sqrt{3}}{2} & \frac{\sqrt{3}}{2} & 0 \\
-z &  -z & z 
\rmatrix,
\eea
with $z\neq0$ and $|E^1| = |E^2| = |E^3|$. The orthonormal 3-torus corresponds to $z=1/\sqrt{2}$ while
the smallest finite volume effects occur at $z=\sqrt{2}$ with $\alpha_{max} = 2^{1/6}$, again the largest
possible value.

\subsection{${\mathbb Z}_4^\prime$}

The second topology allowing for $\vartheta = \pi/2$ rotations has spatial metric,
\bea
E = \frac{L}{|z|^{1/3}} \lmatrix{ccc} 
1 & 0 & \frac{1}{2} \\
0 & 1 & \frac{1}{2} \\
0 & 0 & z 
\rmatrix,\nn 
\eea
with $z\neq0$. An orthonormal 3-torus is not possible in this family
and the one with the smallest finite volume effects occur at $z=1/\sqrt{2}$ with $\alpha_{max} =
2^{1/6}$, again the largest possible value.

\section{Moving frame with finite rotations}

In all 7 cases discussed above, spatial translations in the third direction are isometries of ${\mathbb
T}^4/G$ hence with $y = (0,0,y_3)$ we can describe moving boxes and finite rotations, i.e. partition
functions of the type (\ref{rotmove}). The unitary operator $e^{i y_3 P_3}$ commutes with the Poincare
invariant Hamiltonian as well as with $U(\vartheta)$ corresponding to rotations by $\boldsymbol R$. The
number of such directions is directly related to the topology of ${\mathbb T}^4/G$. Namely if $B$ is the
first Betti number of ${\mathbb T}^4/G$, i.e.  we have the integral cohomology $H^1({\mathbb
T}^4/G,{\mathbb Z}) = {\mathbb Z}^B$ then the number of spatial translations which commute with the
Hamiltonian and $U(\vartheta)$ is $B-1$. This is because one direction in all 7 cases corresponds to time
translation.  Now it is easy to see that $B=2$ hence there is a single spatial direction of interest and
it is along the $z$-direction.

This means we can add shifted boundary conditions to (\ref{rrr}) and (\ref{xprime}). 
The spatial boundary conditions
remain unchanged, the temporal one is modified to,
\bea
\phi\left(t+\frac{L_0}{k},x_1,x_2,x_3\right) & = & \phi(t,x_1^\prime,x_2^\prime,x_3-y_3) \nn
\eea
and the identification of temperature is modified as well, $\sqrt{L_0^2/k^2+y_3^2} = 1/T$, assuming
$-L_3/2 < y_3 < L_3/2$ which can always be achieved.

\section{Further compact, flat, orientable manifolds}

Beyond the 7 topologies considered in the previous sections corresponding to an abelian group, we have
identified 9 further examples allowing for a similar interpretation.
The corresponding groups $G$ may be obtained by adding another generator $(A_2,b_2)$ 
such that $A_2$ rotates and $b_2$ shifts spatially only. As far as the rotation and shift contained in
the original generator $(A,b)$
is concerned the setup is unchanged from the previous sections but the new generator will
mean that we have rotated boundary conditions also spatially. Hence the spatial volume can be thought of
as ${\mathbb T}^3/H$ where the group $H$ is the one generated by $(A_2,b_2)$ and can be again 
one of ${\mathbb Z}_{2,3,4,6}$. Now the original abelian 
group, generated by $(A,b)$, and $H$, generated by $(A_2,h_2)$, may combine into an abelian or
non-abelian group $G$, depending on whether $A$ and $A_2$ commute or not. If the former, we will have
$G = {\mathbb Z}_2 \times {\mathbb Z}_2$, if the latter, $G$ will be one of the following, 
$D_6 = S_3, A_4, D_8, D_{12}$, using standard notation for the symmetric, dihedral and alternating
groups.

The remaining non-trivial compact, flat, orientable spaces are 
still perfectly well-defined compactifications from ${\mathbb R}^4$ and may have applications in QFT.
We will return to QFT on these novel Euclidean space-times in future publications.

\section{Conclusion and outlook}

We have discussed, in the continuum, the path integral formulation of a spatially periodic system at finite
temperature involving finite rotations by a constant angular velocity $\omega$. 
In Euclidean signature $\omega$ is purely imaginary and corresponds to a real angle $\vartheta = i \omega
/ T$. Since a compact space-time breaks $SO(4)$ not all angles are allowed. We have shown that a
consistent description can be given for $\vartheta = 2\pi/k$ with $k=2,3,4,6$ only, in terms of
space-time manifolds ${\mathbb T}^4 / {\mathbb Z}_k$ and these all are flat, orientable and
without boundary. These properties were the main reason for our study motivated by lattice
field theory.
Furthermore for $k=2,3,4$ we have two topologically 
distinct space-times, leading to a total of 7 options. All of these can be realized by fields subject
to novel rotated boundary conditions, opening the way to a path integral representation.

Our discussion naturally led to QFT on compact, flat, orientable manifolds and may of course involve
fermions. For the latter a spin structure is needed and only 23 of the 26 manifolds, including the 7 we
have studied, possess a spin structure \cite{math1,math2,math3}
We will return to gauge theories coupled to fermions in future publications.

As emphasized throughout, we are working on Euclidean space-times and the question of Wick rotating back
to Minkowski signature is an important one. In the Euclidean setup there does not appear to be a
limitation on the spatial volume, it can be arbitrarily large. But clearly on the Minkowski side the
spatial volume cannot be too large, because, as mentioned in section \ref{introduction}, a rigid rotation
of a too large spatial volume will lead to points moving faster than the speed of light. Hence clearly a
straightforward Wick rotation from Euclidean to Minkowski signature can only be expected to hold for not
too large volumes. This subtlety is of course also present for the rotating frame approaches
when the Lagrangian is modified locally by an angular velocity dependent metric and the temporal boundary conditions 
are the usual ones and the spatial ones are those compatible with infinitesimal rotations. 

A further note
about the thermodynamic limit might be useful. As mentioned, there does not appear to be a limition on the
spatial volume in Euclidean signature. The thermodynamic limit, at finite temperature, 
can be taken and the effect of a finite
rotation angle is expected to persist. This is because the non-trivial boundary condition is in the
temporal direction and at finite temperature the temporal extent stays finite of course. At zero
temperature, however, the question is much more delicate. Our construction is explicitly at non-zero
temperature and since the angles are the (imaginary) angular velocity in temperature units, the zero
temperature limit is far from being clear and deserve to be studied in more detail in the future.

Since our spatial 3-torus boundary conditions are only compatible with finite rotations and
the corresponding operator $U(\vartheta)$ is not the exponential of a local angular momentum operator $J$, 
it is worthwhile to point out again the main differences and limitations
relative to the rotating frame approach with
spatial boundary conditions allowing for infinitesimal rotations.
In the latter, higher moments and cumulants of $J$ can be defined without problems. In the former, our
setup, this is not the
case simply because $J$ does not exist as a local operator and also because there is not an arbitrary
angle parameter with respect to which derivatives could be taken. Rotations are at fixed finite angles. 
What can be defined without problems though
is the expectation values of the operator $U(\vartheta)$ and its various powers and their deviations from unity. 

This state of affairs may be constrasted with the moving frame setup 
\cite{Giusti:2010bb,Giusti:2011kt,Giusti:2012yj,Giusti:2014ila,Giusti:2016iqr}
where also local velocity-dependent changes in the action may 
be traded for a temporal non-trivial shifted boundary condition. In this case,
however, the boundary condition makes sense for arbitrary shifts and derivatives with respect to the
shift vector $y$ can be defined. Hence higher moments and cumulants of the corresponding momentum operator $P$ 
can also be defined and compared between the two approaches: local changes to the action and usual
temporal boundary conditions vs. no change in the action and shifted temporal boundary conditions. The
two should of course agree.

A similar comparison for rotations can meaningfully be done on a spatial cylinder
where infinitesimal rotations do make sense both in the rotating frame approach and the rotated temporal
boundary condition approach. Derivatives with respect to the angle provide higher moments and
cumulants of $J$ in both cases, and they of course agree. With our spatial 3-torus boundary conditions,
corresponding to a finite box, the limitation is that the rotating frame approach cannot be set up, only
the rotated temporal boundary conditions and higher moments and cumulants of $J$ cannot be defined
directly.

Naturally, all of our discussions were in the continuum. Discretization on a lattice does not seem to
present conceptual problems and the explicit boundary conditions we have derived can be straightforwardly
implemented in numerical simulations. Since most lattice field theory simulations are performed on a
discretization of ${\mathbb T}^4$, the modifications necessary are rather minor. This also means that a
comparison with simulation results at zero rotation are straightforward.

In conclusion, we have presented a new formulation of Euclidean QFT on flat, compact manifolds without
boundary that we believe could be helpful in studying rotating systems and complement previous
approaches. Important questions remain however as to its interpretation and their precise comparisons.

\vspace{0.5cm}

\begin{acknowledgments}
DN would like to thank Rafa\l$\,$ Lutowski for numerous very enlightening discussions.
This work was supported by NKFIH grants No. TKP2021-NKTA-64, NKKP Excellence No. 151482 and K-147396. 
\end{acknowledgments}


\begin{thebibliography}{99}


\bibitem{Vilenkin:1979ui}
A.~Vilenkin,
Phys. Rev. D \textbf{20}, 1807-1812 (1979)

\bibitem{Erdmenger:2008rm}
J.~Erdmenger, M.~Haack, M.~Kaminski and A.~Yarom,
JHEP \textbf{01}, 055 (2009)
[arXiv:0809.2488 [hep-th]].

\bibitem{Banerjee:2008th}
N.~Banerjee, J.~Bhattacharya, S.~Bhattacharyya, S.~Dutta, R.~Loganayagam and P.~Surowka,
JHEP \textbf{01}, 094 (2011)
[arXiv:0809.2596 [hep-th]].

\bibitem{Son:2009tf}
D.~T.~Son and P.~Surowka,
Phys. Rev. Lett. \textbf{103}, 191601 (2009)
[arXiv:0906.5044 [hep-th]].

\bibitem{Kharzeev:2010gr}
D.~E.~Kharzeev and D.~T.~Son,
Phys. Rev. Lett. \textbf{106}, 062301 (2011)
[arXiv:1010.0038 [hep-ph]].

\bibitem{Landsteiner:2011cp}
K.~Landsteiner, E.~Megias and F.~Pena-Benitez,
Phys. Rev. Lett. \textbf{107}, 021601 (2011)
[arXiv:1103.5006 [hep-ph]].

\bibitem{Kharzeev:2015znc}
D.~E.~Kharzeev, J.~Liao, S.~A.~Voloshin and G.~Wang,
Prog. Part. Nucl. Phys. \textbf{88}, 1-28 (2016)
[arXiv:1511.04050 [hep-ph]].

\bibitem{Chen:2015hfc}
H.~L.~Chen, K.~Fukushima, X.~G.~Huang and K.~Mameda,
Phys. Rev. D \textbf{93}, no.10, 104052 (2016)
[arXiv:1512.08974 [hep-ph]].

\bibitem{Ebihara:2016fwa}
S.~Ebihara, K.~Fukushima and K.~Mameda,
Phys. Lett. B \textbf{764}, 94-99 (2017)
[arXiv:1608.00336 [hep-ph]].

\bibitem{Fukushima:2018grm}
K.~Fukushima,
Prog. Part. Nucl. Phys. \textbf{107}, 167-199 (2019)
[arXiv:1812.08886 [hep-ph]].

\bibitem{Fukushima:2020ncb}
K.~Fukushima, T.~Shimazaki and L.~Wang,
Phys. Rev. D \textbf{102}, no.1, 014045 (2020)
[arXiv:2004.05852 [hep-ph]].

\bibitem{Chen:2022smf}
S.~Chen, K.~Fukushima and Y.~Shimada,
Phys. Rev. Lett. \textbf{129}, no.24, 242002 (2022)
[arXiv:2207.12665 [hep-ph]].

\bibitem{Chernodub:2022veq}
M.~N.~Chernodub, V.~A.~Goy and A.~V.~Molochkov,
Phys. Rev. D \textbf{107}, no.11, 114502 (2023)
[arXiv:2209.15534 [hep-lat]].

\bibitem{Chernodub:2022qlz}
M.~N.~Chernodub,
[arXiv:2210.05651 [quant-ph]].

\bibitem{Chen:2024tkr}
S.~Chen, K.~Fukushima and Y.~Shimada,
Phys. Lett. B \textbf{859}, 139107 (2024)
[arXiv:2404.00965 [hep-ph]].

\bibitem{Fukushima:2025hmh}
K.~Fukushima and Y.~Shimada,
Phys. Lett. B \textbf{868}, 139716 (2025)
[arXiv:2506.03560 [hep-ph]].

\bibitem{Grenier:2015pya}
I.~A.~Grenier and A.~K.~Harding,
Comptes Rendus Physique \textbf{16}, 641-660 (2015)
[arXiv:1509.08823 [astro-ph.HE]].

\bibitem{Watts:2016uzu}
A.~L.~Watts, \textit{et al.}
Rev. Mod. Phys. \textbf{88}, no.2, 021001 (2016)
[arXiv:1602.01081 [astro-ph.HE]].

\bibitem{Paschalidis:2016vmz}
V.~Paschalidis and N.~Stergioulas,
Living Rev. Rel. \textbf{20}, no.1, 7 (2017)
[arXiv:1612.03050 [astro-ph.HE]].

\bibitem{Becattini:2007sr}
F.~Becattini, F.~Piccinini and J.~Rizzo,
Phys. Rev. C \textbf{77}, 024906 (2008)
[arXiv:0711.1253 [nucl-th]].

\bibitem{Jiang:2016woz}
Y.~Jiang, Z.~W.~Lin and J.~Liao,
Phys. Rev. C \textbf{94}, no.4, 044910 (2016)
[erratum: Phys. Rev. C \textbf{95}, no.4, 049904 (2017)]
[arXiv:1602.06580 [hep-ph]].

\bibitem{STAR:2017ckg}
L.~Adamczyk \textit{et al.} [STAR],
Nature \textbf{548}, 62-65 (2017)
[arXiv:1701.06657 [nucl-ex]].

\bibitem{Ambrus:2014uqa}
V.~E.~Ambru{\c{s}} and E.~Winstanley,
Phys. Lett. B \textbf{734}, 296-301 (2014)
[arXiv:1401.6388 [hep-th]].

\bibitem{Chernodub:2016kxh}
M.~N.~Chernodub and S.~Gongyo,
JHEP \textbf{01}, 136 (2017)
[arXiv:1611.02598 [hep-th]].

\bibitem{Yamamoto:2013zwa}
A.~Yamamoto and Y.~Hirono,
Phys. Rev. Lett. \textbf{111}, 081601 (2013)
[arXiv:1303.6292 [hep-lat]].

\bibitem{Chernodub:2017ref}
M.~N.~Chernodub and S.~Gongyo,
Phys. Rev. D \textbf{95}, no.9, 096006 (2017)
[arXiv:1702.08266 [hep-th]].

\bibitem{Braguta:2021jgn}
V.~V.~Braguta, A.~Y.~Kotov, D.~D.~Kuznedelev and A.~A.~Roenko,
Phys. Rev. D \textbf{103}, no.9, 094515 (2021)
[arXiv:2102.05084 [hep-lat]].

\bibitem{Braguta:2023iyx}
V.~V.~Braguta, M.~N.~Chernodub and A.~A.~Roenko,
Phys. Lett. B \textbf{855}, 138783 (2024)
[arXiv:2312.13994 [hep-lat]].

\bibitem{Braguta:2023yjn}
V.~V.~Braguta, M.~N.~Chernodub, A.~A.~Roenko and D.~A.~Sychev,
Phys. Lett. B \textbf{852}, 138604 (2024)
[arXiv:2303.03147 [hep-lat]].

\bibitem{Braguta:2023tqz}
V.~V.~Braguta, M.~N.~Chernodub, I.~E.~Kudrov, A.~A.~Roenko and D.~A.~Sychev,
Phys. Rev. D \textbf{110}, no.1, 014511 (2024)
[arXiv:2310.16036 [hep-ph]].

\bibitem{Giusti:2010bb}
L.~Giusti and H.~B.~Meyer,
Phys. Rev. Lett. \textbf{106}, 131601 (2011)
[arXiv:1011.2727 [hep-lat]].

\bibitem{Giusti:2011kt}
L.~Giusti and H.~B.~Meyer,
JHEP \textbf{11}, 087 (2011)
[arXiv:1110.3136 [hep-lat]].

\bibitem{Giusti:2012yj}
L.~Giusti and H.~B.~Meyer,
JHEP \textbf{01}, 140 (2013)
[arXiv:1211.6669 [hep-lat]].

\bibitem{Giusti:2014ila}
L.~Giusti and M.~Pepe,
Phys. Rev. Lett. \textbf{113}, 031601 (2014)
[arXiv:1403.0360 [hep-lat]].

\bibitem{Giusti:2016iqr}
L.~Giusti and M.~Pepe,
Phys. Lett. B \textbf{769}, 385-390 (2017)
[arXiv:1612.00265 [hep-lat]].

\bibitem{bieberbach1}
L.~Bieberbach,
Math. Ann. \textbf{70}, 297 (1911)

\bibitem{bieberbach2}
L.~Bieberbach,
Math. Ann. \textbf{72}, 400 (1912)

\bibitem{hatcher}
A.~Hatcher, Algebraic Topology, Cambridge University Press (2002)

\bibitem{Ratcliffe:2006bfa}
J.~G.~Ratcliffe,
ISBN 978-0-387-47322-2, 
Springer (2006) 

\bibitem{book}
A.~Szczepański,
Algebra and Discrete Mathematics, Vol 4, World Scientific Publishing Company (2012)

\bibitem{Luscher:1985dn}
M.~Luscher,
Commun. Math. Phys. \textbf{104}, 177 (1986)

\bibitem{Gasser:1986vb}
J.~Gasser and H.~Leutwyler,
Phys. Lett. B \textbf{184}, 83-88 (1987)

\bibitem{Lenstra:1982eee}
A.~K.~Lenstra, H.~W.~Lenstra and L.~Lov{\'a}sz,
Math. Ann. \textbf{261}, no.4, 515-534 (1982)

\bibitem{math1}
B.~Putrycz and A.~Szczepanski, 
Adv. in Geometry, 10 (2), 323-332 (2010)

\bibitem{math2}
R.~Lutowski and B.~Putrycz, 
J. Algebra 436, 277-291 (2015)

\bibitem{math3}
R.~Lutowski, N.~Petrosyan and A.~Szczepański, 
Mathematika, Vol. 64(1), 253-266 (2018)






\end{thebibliography}
\end{document}